\documentclass[runningheads]{llncs}
\usepackage{ecir25-longeval}

\begin{document}
\title{Counterfactual Query Rewriting \\ to Use Historical Relevance Feedback}

\author{
    Jüri Keller\inst{1} \orcidID{0000-0002-9392-8646} \and
    Maik Fr{\"{o}}be\inst{2} \orcidID{0000-0002-1003-981X} \and
    Gijs Hendriksen\inst{3} \orcidID{0000-0003-0945-3148} \and
    Daria Alexander\inst{3} \orcidID{0000-0001-9478-7083} \and
    Martin Potthast\inst{4} \orcidID{0000-0003-2451-0665} \and
    Matthias Hagen\inst{2} \orcidID{0000-0002-9733-2890} \and
    Philipp Schaer\inst{1} \orcidID{0000-0002-8817-4632} 
}

\authorrunning{Keller et al.}

\institute{
    TH Köln - University of Applied Sciences, Cologne, Germany \and
    Friedrich-Schiller-Universität Jena \and
    Radboud Universiteit Nijmegen \and
    University of Kassel, hessian.AI, ScaDS.AI
}

\maketitle

\begin{abstract}
    When a retrieval system receives a query it has encountered before, previous relevance feedback, such as clicks or explicit judgments can help to improve retrieval results. However, the content of a previously relevant document may have changed, or the document might not be available anymore. Despite this evolved corpus, we counterfactually use these previously relevant documents as relevance signals. In this paper we proposed approaches to rewrite user queries and compare them against a system that directly uses the previous qrels for the ranking. We expand queries with terms extracted from the previously relevant documents or derive so-called keyqueries that rank the previously relevant documents to the top of the current corpus. Our evaluation in the CLEF LongEval scenario shows that rewriting queries with historical relevance feedback improves the retrieval effectiveness and even outperforms computationally expensive transformer-based approaches.
\end{abstract}


\section{Introduction}

Many queries received by a search engine have been seen before~\cite{DBLP:journals/sigir/SilversteinHMM99}. Analyzing how users interact with the results of these queries provides valuable relevance indicators that can improve the search engine's effectiveness, e.g., by constructing click models that synthesize a relevance indicator~\cite{chuklin:2015}. These synthesized labels can be used to boost documents~\cite{keller:2024b}, to train learning-to-rank models~\cite{liu:2011}, or to fine-tune transformer models~\cite{lin:2021}. While the training and fine-tuning of models often require huge amounts of labeled data, boosting also works with few labels~\cite{keller:2024b}. However, it may only be effective shortly after relevance information has been observed, as queries, documents, and relevance may evolve~\cite{keller:2024}. For instance, boosting would fail when documents are deleted or change their content such that they are not relevant to the query anymore (exemplified in Table~\ref{table-examples}).

\begin{table}[t]
    \centering
    \renewcommand{\tabcolsep}{4pt}
    \caption{Historical relevance feedback observed at timestamp~$t_{0}$ for a query $q := $~\texttt{bird song} that is to be applied at timestamp~$t_{1}$. Transferring $t_{0}$ to $t_{1}$ would introduce errors, whereas our counterfactual query rewriting always uses the correct $t_{0}$ observations.}
    \label{table-examples}
    \vspace*{-2ex}
    \begin{tabular}[t]{@{}lllc@{}}
    \toprule
   \multicolumn{2}{@{}c@{}}{\bfseries Document} & {\bfseries Comment} & {\bfseries Type}\\
    \cmidrule(r@{\tabcolsep}){1-2}
    
    Timestamp~$t_0$ &  Timestamp~$t_1$\\
    \midrule

    \multirow{2}{*}{ \shortstack[l]{\\ Alphabetical list of {\color{blue} \emph{bird}} \\ {\color{blue} \emph{songs}} you may like \ldots\vspace*{-.1cm}}} & \multirow{2}{*}{ \shortstack[l]{~~~~~~~~~---}} & \multirow{2}{*}{ \shortstack[l]{Document deleted \\ from the web}}  &   \multirow{2}{*}{ \shortstack[l]{\texttt{DELETE}}}\\\\
    \midrule

    \multirow{2}{*}{ \shortstack[l]{\\ Best phone ring tone?\\ Enjoy {\color{blue} \emph{bird songs}} \ldots\vspace*{-.1cm}}} & \multirow{2}{*}{ \shortstack[l]{\\ Get phone tones from the\\ charts for free \ldots\vspace*{-.1cm}}}  & \multirow{2}{*}{ \shortstack[l]{Document became \\ non-relevant}}  & \multirow{2}{*}{ \shortstack[l]{\texttt{UPDATE}}}\\\\

    \midrule

    \multirow{2}{*}{ \shortstack[l]{\\ 311~{\color{blue} \emph{songs by birds}} from\\ France by species \ldots\vspace*{-.1cm}}} & \multirow{2}{*}{ \shortstack[l]{\\ 312~{\color{blue} \emph{songs by birds}} from\\ France by species \ldots\vspace*{-.1cm}}}  & \multirow{2}{*}{ \shortstack[l]{Document remains \\ relevant}}  & \multirow{2}{*}{ \shortstack[l]{\texttt{UPDATE}}}\\\\

    \bottomrule
\end{tabular}
\vspace*{-1ex}
\end{table}

To address this challenge, we explore how historical relevance feedback can guide query rewriting approaches. Incorporating the historical relevance feedback into query rewriting works, analogously to boosting, already with few feedback documents~(e.g., RM3~in PyTerrier uses 3~feedback documents as default~\cite{macdonald:2020}). The document corpus naturally evolves over time for various reasons, such as newly created content, updated websites, or the removal of outdated information. While the conventional retrieval setting relies on the factual, current state of the collection. In contrast, we additionally use previous versions of the collection and thereby counterfactually assume that superseded documents remain relevant or are at least suitable for constructing relevance indicators. This counterfactual assumption is motivated by the observation that many document changes are incremental and may not substantially alter their relevance to the original query.

We apply our counterfactual relevance feedback in three scenarios: (1)~via boosting, (2)~via explicit relevance feedback, and (3)~via keyqueries. Boosting re-weights known (non-)relevant documents, which can not generalize to new or deleted documents. Explicit relevance feedback extends the query with terms from known relevant documents but does not test the resulting ranking. Therefore, we use so-called keyqueries~\cite{gollub:2013a} which reformulate queries until the target documents appear in the top positions.

We evaluate our counterfactual query rewriting approaches on the LongEval test collection. Our results show that counterfactual query rewriting is as effective as boosting but generalizes to new documents,  making it more robust than boosting and substantially outperforming neural transformers while being much more efficient as queries can be pre-computed. Our code is publicly available.%
\footnote{\url{https://github.com/webis-de/ECIR-25}}

\section{Related Work}
\label{sec:related-work}

\paragraph{Web Dynamics.} Temporal dynamics in web search are an established research topic. Websites change constantly, often more than hourly~\cite{DBLP:conf/wsdm/AdarTDE09}, making them relevant for only a limited time~\cite{DBLP:conf/sigir/TikhonovBBOKG13}. This relates directly to the observation that many queries are not unique but frequently reissued~\cite{DBLP:conf/sigir/Dumais14,DBLP:journals/sigir/SilversteinHMM99}. Even the same users tend to repeat the same queries at different points in time~\cite{DBLP:conf/wsdm/TylerT10}.

\paragraph{Temporal Information Retrieval.} The observed dynamics motivate Temporal Information Retrieval~(TIR) aiming to use temporal information to improve the ranking quality~\cite{DBLP:journals/ftir/KanhabuaBN15,DBLP:journals/csur/CamposDJJ14}, e.g., by using temporal patterns for the term weighting~\cite{DBLP:conf/wsdm/ElsasD10}. While TIR focuses on leveraging temporal properties, our work takes a complementary approach by exploiting past document versions rather than directly addressing their temporal aspects.

\paragraph{Query Rewriting with Keyqueries.} Given a set of target documents, a \emph{keyquery} is the minimal query that retrieves the target documents in the top positions~\cite{gollub:2013a,hagen:2016b}. We adapt this approach to the web search setting by using previously relevant documents as target documents. Keyqueries use terms generated via RM3 or other query expansion approaches as vocabulary for an efficient enumeration of query candidates~\cite{froebe:2022c,froebe:2021c}. Beyond other query expansion approaches, the keyquery approach also generates a ranking for each candidate to test if all criteria are fulfilled and thereby fully leveraging historical data. 

\paragraph{Evaluations in Dynamic Settings.} Although temporal dynamics can strongly influence the effectiveness of IR systems, they are rarely  considered during evaluation. Soboroff~\cite{DBLP:conf/sigir/Soboroff06} initially investigated how temporal dynamics influence test collection evaluations and hypothesized how they could be maintained. Fr{\"o}be et al.~\cite{froebe:2022d} studied the case when relevance judgments are re-used between different snapshots of crawled documents. Recently, the LongEval shared task~\cite{alkhalifa:2023,DBLP:conf/clef/AlkhalifaBDEAFG24} provides a test bed of an evolving web search scenario covering over a year. Changes in evolving test collections are described through create, update, and delete operations on documents, topics, and relevance judgments~\cite{keller:2024}. While the LongEval dataset contains changes in all types of components, this study is mainly concerned with changing documents and ignore, counterfactually, changes in the relevance label.

\section{Query Rewriting on Historical Relevance Feedback}

We present three approaches that incorporate historical relevance feedback, from \Ni boosting (can not generalize), over \Nii relevance feedback (might underfit), towards \Niii keyqueries (trade-off under- vs. overfitting). These approaches were previously applied in diverse retrieval scenarios (see Section~\ref{sec:related-work}). However, we are the first to adapt them to a retrieval scenario that evolves over time.

For a set $H$~of historical relevance feedback with observations $(q, d, t) \in H$, $rel(q, d, t)$ is the (graded) relevance judgment for document~$d$ for query~$q$ at timestamp~$t$. Document~$d$ may have been deleted or substantially changed in the current corpus~(Table~\ref{table-examples}), which is why we counterfactually use the version of the document at the timestamp $t$ where the relevance observation was made.

\paragraph{Boosting Previously Relevant Documents.} Given a ranking~$r$ and a set of historical relevance feedback~$H$, boosting incorporates a document's historical relevance label, regardless of how much the document has changed or how old the relevance feedback is. We apply boosting to adjust the scores of documents based on their historical relevance. For a document~$d$ previously observed for the query~$q$ at the timestamps $t_{1}, \ldots, t_{k}$, we increase the score if it is relevant, respectively decrease the score if it is not-relevant at the corresponding timestamps using a weighting factor $\lambda^2$ and additionally boost highly relevant documents by a factor of $\mu$. The boosted score of document $d$ for query $q$ is then computed by:

\begin{equation}
    \text{score}(q,d) \; = \; \text{score}_0 \;\times \;
    \prod_{t = t_1}^{t_k}
    \begin{cases}
        (1 - \lambda)^2, & \text{if } rel(q,d,t) = 0, \\
        \lambda^2, & \text{if } rel(q,d,t) = 1, \\
        \lambda^2  \mu, & \text{if } rel(q,d,t) = 2.
        \end{cases}
\end{equation}

While this qrel boosting is highly effective when documents do not change~\cite{alkhalifa:2024,keller:2024b}, it cannot generalize to newly created or deleted documents.

\paragraph{Previously Relevant Documents as Relevance Feedback.} Given a retrieval model, the current document corpus~$D$, and historical relevance feedback~$H$, we expand each query by adding $k$~terms with the highest tf-idf scores of previously known relevant documents. For a query $q$, the set $D^{+} = \{d| (q,d,t) \in H \wedge rel(q,d,t) > 0\}$ specifies the previously positive documents on which we calculate the tf-idf scores. The top $k$~terms with the highest tf-idf scores are obtained for query expansion and appended to the original query. The expanded query is submitted to the retrieval system on the current document corpus $D$ to produce the final ranking. Since this expansion relies solely on tf-idf scores from previous corpora, these scores can be calculated offline. Improving upon boosting, this allows to generalize from the historical relevance feedback to potentially deleted or newly created documents.

\paragraph{Keyqueries for Previously Relevant Documents.} Given a retrieval model, the current document corpus~$D$, and a set of historical relevance feedback~$H$, we construct keyqueries~\cite{froebe:2021c,gollub:2013a,hagen:2016b} against the previously relevant documents. A query $q_{k}$ is a keyquery for the set of target documents $D^{+} = \{d| (q,d,t) \in H \wedge rel(q,d,t) > 0\}$ previously known relevant for a query~$q$ against the corpus $D^{+} \cup D$ for the given retrieval model, iff \Ni every $d \in D^{+}$ is in the top-$k$ results, \Nii $q_{k}$ has more than $l$ results, and \Niii no subquery $q^{'}_{k} \subset q_{k}$ satisfies the above. The first criterion ensures the specificity, the second the generality, and the third the minimality of the keyquery, together trading off generalizability versus specificity. Traditional relevance feedback does not verify the position of feedback documents in the resulting rankings, whereas keyqueries uses them to remove overfitted or underfitted candidates. To generate candidates, we re-implemented a previous algorithm~\cite{froebe:2022c} in PyTerrier~\cite{macdonald:2020} which generates candidates from the top-10 RM3 terms. If multiple candidates are keyqueries, we use the one with the highest nDCG@10 on $D^{+}$. Finally, the keyquery is submitted against the corpus $D$, which may, or may not, contain documents that were previously relevant.

\section{Evaluation}

We evaluate our counterfactual query rewriting in the LongEval scenario that comes with overlapping queries across six points in time between June~2022 and August~2023~\cite{alkhalifa:2023,alkhalifa:2024,galuscakova:2023}. We modify the LongEval datasets to focus on queries that re-occur across multiple timestamps to study the effects of evolving documents. We evaluate the retrieval effectiveness of all approaches and use an ablation study to investigate if they generalize beyond previously known relevant documents.

\subsection{Experimental Setup}
\begin{figure}[t]
    \begin{minipage}{.48\textwidth}
        \includegraphics[width=\textwidth]{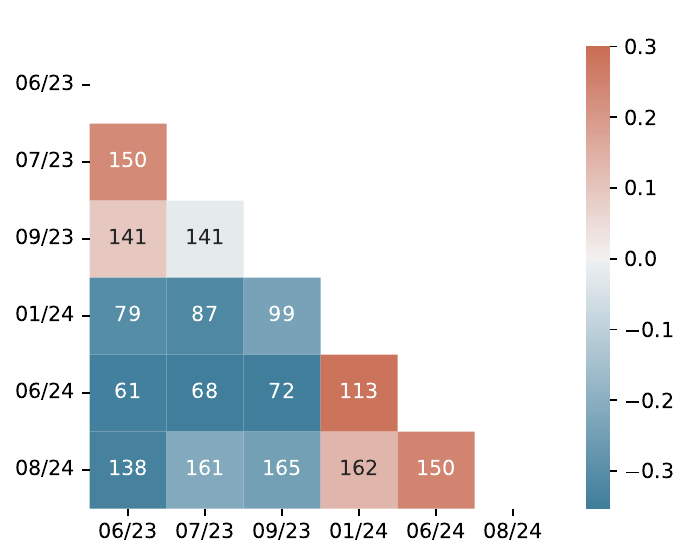}
        \vspace{-4ex}
        \caption{Frequency of overlapping queries over the different timestamps.}
        \label{fig:query-overlap}
    \end{minipage}
    \hfill    
    \begin{minipage}{.50\textwidth}
        \includegraphics[width=\textwidth]{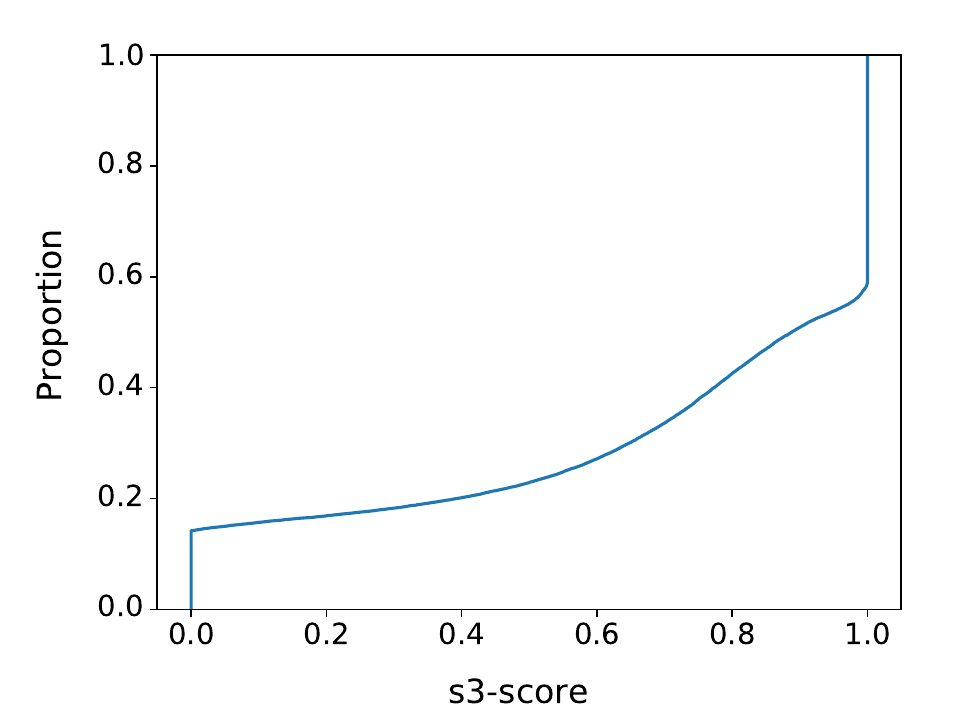}
        \vspace{-4ex}
        \caption{S$_{3}$ Similarities of documents with overlapping URLs as eCDF plot.}
        \label{fig:document-similarities}
    \end{minipage}
\end{figure}

The LongEval test collection~\cite{galuscakova:2023} samples documents, queries, and clicks from the French web search engine Qwant. For each timestamp, we remove queries that did not occur at least in one earlier timestamp, leaving 5~time\-stamps for evaluation between July~2022 and August~2023. Figure~\ref{fig:query-overlap} overviews the overlapping queries for the timestamps. For instance, 138~queries from June~2023 re-occur in August~2024, forming the biggest time gap in our evaluation scenario.

We contrast five baselines with our three approaches. We use BM25~\cite{robertson:1994}, BM25 with RM3 expansion (implemented in PyTerrier~\cite{macdonald:2020}), ColBERT~\cite{khattab:2020}, List-in-T5~\cite{tamber:2023}, and monoT5~\cite{nogueira:2020}. We use the default hyperparameters for all baselines (exporting ColBERT, List-in-T5, and monoT5 from TIRA/TIREx~\cite{froebe:2023e,froebe:2023b}). We also implement our three approaches in PyTerrier using BM25 as the underlying retrieval model. For boosting (BM25$_{Boost}$), we set $\lambda=0.7$ and $\mu=2$ based on previous experiments~\cite{keller:2024b}. For relevance feedback (BM25$_{RF}$), $k=10$ feedback terms are used as this is also the default for RM3 in PyTerrier. For keyqueries (BM25$_{keyquery}$), we use 10~feedback terms aiming at queries that retrieve the target tocuments to the top-10 while having more than 25~results.

\subsection{Evolution of Documents in the LongEval Corpora over Time}

The documents in the corpora may evolve via deletion, creation, or updates. The corpus comprises between one and 2.5~million documents, with a total 2.6~million created and 1.7~million deleted over time. We measure how the re-occuring documents changed by inspecting their pairwise similarities. We measured the similarity with the $S_{3}$ score~\cite{bernstein:2005} implemented in CopyCat~\cite{froebe:2021a} at default configuration (1~indicates identical, 0~no overlap) as this score aims to identify redundant documents in retrieval scenarios~\cite{bernstein:2005}. Figure~\ref{fig:document-similarities} shows the $S_{3}$ similarities for all document pairs, indicating that 40\,\% do not change their content~($S_{3}=1.0$), whereas around 50\,\% have an $S_{3}$ similarity below~0.8 that indicate non-negligible changes (prior research used 0.82~as near-duplicate threshold on the Web~\cite{froebe:2021a}). Given that the LongEval corpora evolve only slightly, we include an ablation study that removes all overlap to analyze how approaches generalize.

\subsection{Retrieval Effectiveness}
\begin{table}[t]
\small
\centering
\renewcommand{\tabcolsep}{1.6pt}
\caption{Retrieval effectiveness of the five baselines and our three approaches as nDCG@10 with and without unjudged documents (nDCG@10$^{'}$) on the LongEval timestamps. We report Bonferroni corrected significance against BM25 ($\dagger$) and monoT5 ($\ddagger$).}
\label{tab:table-results}

\begin{tabular}{@{}l@{}cccccccccc@{}}
    \toprule
    \bfseries System & \multicolumn{5}{c}{\bfseries nDCG@10} & \multicolumn{5}{c}{\bfseries nDCG@10$^{'}$}\\
    \cmidrule(r@{.25em}){2-6}
    \cmidrule(l@{.25em}){7-11}
    
    & 07/22 & 09/22 & 01/23 & 06/23 & 08/23 & 07/22 & 09/22 & 01/23 & 06/23 & 08/23\\
    
    \midrule

BM25 & .155$^{\phantom{\dagger\ddagger}}$ & .184$^{\phantom{\dagger\ddagger}}$ & .172$^{\phantom{\dagger\ddagger}}$ & .175$^{\phantom{\dagger\ddagger}}$ & .134$^{\phantom{\dagger\ddagger}}$ & .471$^{\phantom{\dagger\ddagger}}$ & .492$^{\ddagger\phantom{\dagger}}$ & .516$^{\ddagger\phantom{\dagger}}$ & .486$^{\ddagger\phantom{\dagger}}$ & .379$^{\ddagger\phantom{\dagger}}$ \\
    BM25$_{RM3}$ & .147$^{\ddagger\phantom{\dagger}}$ & .181$^{\phantom{\dagger\ddagger}}$ & .163$^{\phantom{\dagger\ddagger}}$ & .174$^{\phantom{\dagger\ddagger}}$ & .134$^{\phantom{\dagger\ddagger}}$ & .478$^{\ddagger\phantom{\dagger}}$ & .490$^{\ddagger\phantom{\dagger}}$ & .524$^{\ddagger\phantom{\dagger}}$ & .492$^{\ddagger\phantom{\dagger}}$ & .388$^{\ddagger\phantom{\dagger}}$ \\
    ColBERT & .198$^{\phantom{\dagger\ddagger}}$ & .207$^{\phantom{\dagger\ddagger}}$ & .201$^{\phantom{\dagger\ddagger}}$ & .184$^{\phantom{\dagger\ddagger}}$ & .151$^{\phantom{\dagger\ddagger}}$ & .402$^{\dagger\phantom{\ddagger}}$ & .409$^{\dagger\phantom{\ddagger}}$ & .420$^{\dagger\phantom{\ddagger}}$ & .408$^{\dagger\phantom{\ddagger}}$ & .315$^{\dagger\phantom{\ddagger}}$ \\
    List-in-T5 & .203$^{\phantom{\dagger\ddagger}}$ & .204$^{\phantom{\dagger\ddagger}}$ & .202$^{\phantom{\dagger\ddagger}}$ & .198$^{\phantom{\dagger\ddagger}}$ & .161$^{\phantom{\dagger\ddagger}}$ & .401$^{\dagger\phantom{\ddagger}}$ & .413$^{\dagger\phantom{\ddagger}}$ & .425$^{\dagger\phantom{\ddagger}}$ & .413$^{\dagger\phantom{\ddagger}}$ & .317$^{\dagger\phantom{\ddagger}}$ \\
    monoT5 & .202$^{\phantom{\dagger\ddagger}}$ & .219$^{\phantom{\dagger\ddagger}}$ & .197$^{\phantom{\dagger\ddagger}}$ & .202$^{\phantom{\dagger\ddagger}}$ & .154$^{\phantom{\dagger\ddagger}}$ & .405$^{\phantom{\dagger\ddagger}}$ & .410$^{\dagger\phantom{\ddagger}}$ & .415$^{\dagger\phantom{\ddagger}}$ & .411$^{\dagger\phantom{\ddagger}}$ & .314$^{\dagger\phantom{\ddagger}}$ \\
    
    \midrule
    
    BM25$_{Boost}$ & \bfseries.355$^{\dagger\ddagger}$ & .372$^{\dagger\ddagger}$ & \bfseries.287$^{\dagger\ddagger}$ & \bfseries.364$^{\dagger\ddagger}$ & \bfseries.271$^{\dagger\ddagger}$ & .529$^{\ddagger\phantom{\dagger}}$ & .546$^{\ddagger\phantom{\dagger}}$ & .541$^{\ddagger\phantom{\dagger}}$ & .540$^{\ddagger\phantom{\dagger}}$ & .412$^{\ddagger\phantom{\dagger}}$ \\
    BM25$_{RF}$ & .303$^{\dagger\ddagger}$ & .332$^{\dagger\ddagger}$ & .241$^{\dagger\phantom{\ddagger}}$ & .262$^{\dagger\ddagger}$ & .191$^{\dagger\ddagger}$ & .606$^{\dagger\ddagger}$ & .611$^{\dagger\ddagger}$ & \bfseries.590$^{\dagger\ddagger}$ & .552$^{\dagger\ddagger}$ & \bfseries .426$^{\dagger\ddagger}$ \\
    BM25$_{keyquery}$ & .350$^{\dagger\ddagger}$ & \bfseries .391$^{\dagger\ddagger}$ & .233$^{\phantom{\dagger\ddagger}}$ & .262$^{\phantom{\dagger\ddagger}}$ & .185$^{\phantom{\dagger\ddagger}}$ & \bfseries.642$^{\phantom{\dagger\ddagger}}$ & .\bfseries 655$^{\ddagger\phantom{\dagger}}$ & .574$^{\dagger\phantom{\ddagger}}$ & \bfseries .554$^{\phantom{\dagger\ddagger}}$ & .422$^{\ddagger\phantom{\dagger}}$ \\

\bottomrule
\end{tabular}
\end{table}

We evaluate the effectiveness of our five baselines and our three approaches using nDCG@10. As the relevance labels of the LongEval corpus are derived from click logs, unjudged documents strongly impact the evaluation. In this scenario, it is recommendet to remove unjudged documents~\cite{sakai:2007} which we report as nDCG@10$^{'}$. Table~\ref{tab:table-results} shows the results. ColBERT, List-in-T5, and monoT5 outperform the BM25 baseline in most cases, whreas BM25 with RM3 expansion does not substantially differ from BM25. Our three approaches substantially outperform all five baselines (nDCG$^{'}$ is always higher). After removing the undesired impact of unjudged documents, both BM25$_{RF}$ and BM25$_{keyquery}$ outperform BM25$_{Boost}$, indicating that these approches generalize to newly created or modified documents. Keyqueries are the most effective approach in all cases, outperforming the best transformer by a large margin.

\begin{table}[t]
\small
\centering
\renewcommand{\tabcolsep}{2.8pt}
\caption{Ablation study showing the nDCG@10$^{'}$ improvement upon BM25  ($\pm$ std-dev) on newly added documents that were never seen before, analyzing how approaches generalize. $^{*}$ marks Bonferroni corrected significance for students t-test.}
\label{tab:table-results-fold}

\begin{tabular}{@{}lcccccccccc@{}}
    \toprule
    \bfseries System & \bfseries 07/23 & \bfseries 09/23 & \bfseries 01/24 & \bfseries 06/24 & \bfseries 08/24 \\
    
    \midrule

BM25$_{Boost}$ & +0.000$^{\phantom{*}}_{\color{gray}\pm.000}$ & +0.000$^{\phantom{*}}_{\color{gray}\pm.000}$ & +0.000$^{\phantom{*}}_{\color{gray}\pm.000}$ & +0.000$^{\phantom{*}}_{\color{gray}\pm.000}$ & +0.000$^{\phantom{*}}_{\color{gray}\pm.000}$ \\
    BM25$_{RF}$ & -0.034$^{\phantom{*}}_{\color{gray}\pm.111}$ & +0.000$^{\phantom{*}}_{\color{gray}\pm.135}$ & +0.022$^{\phantom{*}}_{\color{gray}\pm.146}$ & +0.012$^{\phantom{*}}_{\color{gray}\pm.081}$ & +0.006$^{\phantom{*}}_{\color{gray}\pm.146}$ \\
    BM25$_{keyquery}$ & -0.010$^{\phantom{*}}_{\color{gray}\pm.084}$ & +0.012$^{\phantom{*}}_{\color{gray}\pm.153}$ & +0.032$^{*}_{\color{gray}\pm.105}$ & -0.001$^{\phantom{*}}_{\color{gray}\pm.065}$ & +0.002$^{\phantom{*}}_{\color{gray}\pm.085}$ \\

\bottomrule
\end{tabular}
\end{table}

We conduct an ablation study to analyze if the improvements of BM25$_{RF}$ and BM25$_{keyquery}$ come from a generalization beyound previously known relevant documents. We remove all documents that occur in previous timestamps from the runs and relevance judgments and evaluate nDCG$^{'}$. This way, all remaining documents have never been seen before. Table~\ref{tab:table-results-fold} shows the results as improvement upon BM25 for our three approaches. As BM25$_{Boost}$ can not generalize to new documents, they never improve (improvement is always~$+0.0$). However, both BM25$_{RF}$ and BM25$_{keyquery}$ generalize to unseen documents.

\section{Conclusion and Future Work}
We explored the capabilities of query rewriting approaches for recurring queries. We counterfactually assume that previously relevant documents are still available to use them as explicit relevance feedback.  

The current analysis is subject to several limitations: currently it is restricted to a web search context, investigates a limited time frame, and relies on a single test collection. Moreover, comparing the approaches against systems that continuously learn from the evolving corpus could yield deeper insights. The dynamics of the collection and the ways users interact with the search system strongly influences the effectiveness of the proposed approaches. As collections evolve for various reasons, previously relevant documents may no longer be available for further use. Expiring licenses may prevent the system from using the licensed documents, even if they are not ranked, whereas outdated or shutdown websites may still be available. Similarly, the query types users issue and how the system processes them may be differently well suited. For instance, while short keyword queries, as used in the experiments, may benefit from direct expansion, natural language queries should not be expanded in the same way. While the results indicated improved effectiveness over the observed time frame, the approaches are essentially vulnerable to the Matthews effect, where older documents tend to accumulate exposure. Promoting older documents may create an increased entry barrier for new documents. Addressing such effects and biases over extended time frames is crucial for ensuring a secure and sustainably effective system. 

Our approaches only need a few documents as feedback, and our results show that this already suffices to significantly outperform expensive transformer-based models. The ablation study suggests that the advanced approaches generalize beyond known query-document pairs, making them effective for new documents as well. Interesting directions for future work would be also to handle scenarios where the intent of a query might change, take into account how the documents evolve, or how relevance feedback observed for similar queries can be re-used.

\bibliographystyle{splncs04}
\raggedright
\bibliography{ecir25-longeval-lit}
\end{document}